\documentclass[journal]{IEEEtran}
\usepackage[pdftex]{graphicx}
\usepackage{wrapfig}
\usepackage{subfigure}
\usepackage[symbol]{footmisc}
\usepackage{bm}
\usepackage{bbm}
\usepackage{amsmath}
\usepackage{cases}
\usepackage{color}
\usepackage{outlines}

\usepackage{amssymb}

\usepackage{cite}

\ifCLASSINFOpdf
\else
\fi
\usepackage{amsmath,amssymb,amsthm}

\begin{document}

%
\title{Undirected graphs: is the shift-enabled condition trivial or necessary?}
%
%
%

\author{Liyan~Chen,
Samuel~Cheng$^\ast$,~\IEEEmembership{Senior~Member,~IEEE,}
Kanghang~He,               Lina~Stankovic,~\IEEEmembership{Senior~Member,~IEEE,}
and        ~Vladimir~Stankovic,~\IEEEmembership{Senior~Member,~IEEE} 
\thanks{L. Chen is with the Department
of Computer Science and Technology, Tongji University, Shanghai,
 201804 China and Key Laboratory of Oceanographic Big Data Mining \& Application of Zhejiang Province, Zhejiang Ocean University, Zhoushan, Zhejiang 316022, China (e-mail: chenliyan@tongji.edu.cn).}
\thanks{S. Cheng is with
the School of Electrical and Computer Engineering, University of Oklahoma, OK 74105, USA (email: samuel.cheng@ou.edu).}
\thanks{K. He, L. Stankovic, and V. Stankovic are with Department of Electronic and Electrical Engineering,
         University of Strathclye, Glasgow, G1 1XW U.K.
        (e-mail:\{kanghang.he,~lina.stankovic,~vladimir.stankovic\}@strath.ac.uk).}
\thanks{$^\ast$ Corresponding author.}
\thanks{
}
}

%
%

\markboth{}%
{Shell \MakeLowercase{\textit{et al.}}: Bare Demo of IEEEtran.cls for IEEE Journals}
%



\newtheorem{lemma}{Lemma}

\maketitle

\begin{abstract}
It has recently been shown that, contrary to the wide belief that a shift-enabled condition (necessary for any shift-invariant filter to be representable by a graph shift matrix) can be ignored because any non-shift-enabled matrix can be converted to a shift-enabled matrix, such a conversion in general may not hold for a directed graph with non-symmetric shift matrix. This letter extends this prior work, focusing on undirected graphs where the shift matrix is generally symmetric. We show that while, in this case, the shift matrix can be converted to satisfy the original shift-enabled condition, the converted matrix is not associated with the original graph, that is, it does not capture anymore the structure of the graph signal. We show via a counterexample, that a non-shift-enabled matrix cannot be converted to a shift-enabled one and still maintain the topological structure of the underlying graph,  which is necessary to facilitate localized signal processing.

\end{abstract} %

\begin{IEEEkeywords}
graph signal processing, shift-enabled graphs, shift-invariant filter, undirected graph.
\end{IEEEkeywords}

%
\IEEEpeerreviewmaketitle

\newtheorem{myDef}{Definition}
\newtheorem{Thm}{Theorem}
\newtheorem{remark}{Remark}

\section{Introduction}
\label{sec:intro}
Graph signal processing (GSP) extends classical digital signal processing (DSP) to signals on graphs, and provides a prospective solution to numerous real-world problems that involve signals defined on topologically complicated domains, such as social networks, point clouds, biological networks, environmental and condition monitoring sensor networks \cite{ortega2018graph}. However, there are several challenges in extending classical DSP to signals on graphs, particularly related to the design and application of graph filters.

In classical 1-D DSP, any linear, time-invariant, or shift-invariant, filter that commutes with time shift operator $z^{-1}$ can be represented as a polynomial of $z^{-1}$ leading to $Z$-transform of the filter.
Conversely, if a linear filter can be represented as a polynomial of $z^{-1}$, the filter is linear and shift-invariant. 
Unfortunately, this concept does not simply generalize to GSP, partly because the definition of a ``shift" for a graph is not obvious \cite{Shuman_2013_The_emerging_field}. Commonly, in the GSP literature, a graph is uniquely described by a ``shift'' matrix or a ``shift" operator\footnote{The term ``shift'' comes from the analogy with $z^{-1}$ operator in $Z-$transform of classical DSP.}, $S$~\cite{sandryhaila_2013_discrete,sandryhaila_2014_discrete_frequency,sandryhaila_2014_big_data}, which has been extensively used for time/vertex-domain filter design (see \cite{ortega2018graph}, \cite{Shuman_2013_The_emerging_field} and references therein for frequency-domain and time/vertex-domain filtering). For example, adjacency matrix, for general graphs, and Laplacian matrix, for undirected graphs, are some popular choices for the shift matrix.

In order to make graph filtering feasible, even for very large graphs, it is necessary to perform the filtering operation locally. 
For example, consider a sensor network represented by a graph, where the edges and edge weights of the graph depend on the distance between the sensors, efficient filtering boils down to merely mixing the signals acquired by a sensor with those of the nearest sensors. Otherwise, if the filter output at any graph vertex is a linear combination of inputs at \textit{all} vertices, filtering will be practically infeasible for ``big data" graphs \cite{sandryhaila_2014_big_data}. Therefore, we expect that a node can only impose direct influence to an adjacent node through the shift operator. For practical purposes, it is advantageous to be able to decompose filters in a form of polynomial of such shift matrix.
The importance of this polynomial representation has been reiterated in a recent survey paper (Section II.F of \cite{ortega2018graph}).


Although a nice, but loose, analogy between $S$ and $z^{-1}$ can be established \cite{ortega2018graph}, unlike classical DSP, if a graph filter is shift-invariant (the shift matrix commuting with the target filter), this does not automatically imply that a polynomial representation of the filter exists~\cite{Teke2017Linear}.
Ref. \cite{sandryhaila_2013_discrete} argues that, for any shift matrix $S$, there exists a converted shift matrix $\tilde{S}$ such that graph filter $H$ is a polynomial in $\tilde{S}$. However, it is not sufficient just to have $H$ to be represented as a polynomial of any arbitrary $\tilde S$. \emph{One should also ensure that $\tilde{S} $ indeed describes the same graph as $S$} (see details in Definition \ref{same_structure}), that is, the converted graph shift should keep the same topological structure as the original one.
 

\subsection{Contribution}
It was shown in  \cite{sandryhaila_2013_discrete} that any filter commuting with shift matrix $S$ can be represented as a polynomial in $S$ provided that the characteristic and minimal polynomial of the shift matrix are equal (in the rest of this paper, as in~\cite{Liyanchen2018}, we will refer to this condition as \textit{shift-enabled condition}, see also Definition~\ref{def:shift_enable}). 
However, in  \cite{sandryhaila_2013_discrete}, this condition was immediately disregarded, surmising that one may convert any shift matrix that does not satisfy the shift-enabled condition into one that does. Based on this conclusion, most researchers now always assume that the shift-enabled condition simply holds or ignore the condition completely.
However, it was proved in~\cite{Liyanchen2018}, through a counterexample, that such a conversion may not hold for a directed graph with asymmetric shift matrix. 

In this letter, we focus on  undirected graphs, which have wider applications \cite{Shuman_2013_The_emerging_field}, and illustrate with examples that when the symmetric shift matrix of an undirected graph is non-shift-enabled, the conversion suggested in~\cite{sandryhaila_2013_discrete} could lead to a very different graph that does not necessarily capture the structure of the original graph signal. Namely, though the conversion would provide a shift-enabled graph that facilitates polynomial representation of the shift-invariant filters, the newly designed graph might no longer capture the structure of the graph signal it was originally designed to model\footnote{Note that~\cite{Teke2017Linear} also reiterated the relationship among polynomial representation, shift-invariant, and alias-free filter. However, \cite{Teke2017Linear} did not explicitly investigate the implication of the shift matrix conversion as proposed in~\cite{sandryhaila_2013_discrete}.}, and does not facilitate performing filtering locally. 

Referring to our wireless sensor network example in the introduction,  in the original graph the output of the filtering at each vertex only involves inputs of the vertex's immediate neighborhoods. However,
in the converted graph, sensors that are far apart might be strongly connected, 
that is, each output at a vertex could be a linear combination of inputs at almost all vertices, 
thus filtering in such converted graph will be computationally unaffordable for ``big data'' graphs in practice
which further emphasizes the importance of the shift-enabled condition \cite{Liyanchen2018}.


The outline of the letter is as follows. Section~\uppercase\expandafter{\romannumeral2} describes the basic concepts and key properties of a shift-enabled graph. Section~\ref{sect:counterexample} provides counterexamples to prove that the shift-enabled condition is essential for the symmetric graph. Section~\uppercase\expandafter{\romannumeral4} concludes the letter.

\section{Basic concepts and properties of shift-enabled graphs}

In this section, we briefly review the concepts of shift-enabled graphs and their properties relevant to this letter. For more details, see \cite{sandryhaila_2013_discrete,Shuman_2013_The_emerging_field,sandryhaila_2014_discrete_frequency,sandryhaila_2014_big_data}.

 Let $\mathcal{G}=(V,A)$ be a graph, where $V=\{v_0,v_1,\cdots,v_{n-1}\}$ is a set of vertices and $A\in \mathbb{C}^{n\times n}$ is the adjacency matrix of the graph. 
 Let $\bm{x}=(x_0,x_1,\cdots,x_{n-1})^T$ be a {\em graph signal}, where each sample $x_i \in \bm{x}$ corresponds to a vertex $v_i \in V$. 

In particular, if $\mathcal{G}$ is a directed circular graph, then the corresponding adjacency matrix is given by: 
$A=\begin{pmatrix}
\begin{smallmatrix}
0 & 0 & \cdots & 0 & 1\\
1 & 0 & \cdots & 0 & 0\\
\vdots& \vdots & \ddots & \ddots & \vdots\\
0 & 0 & \cdots & 1 & 0\\
\end{smallmatrix}
\end{pmatrix} 
$. Then
$A\bm{x}=(x_{n-1},x_0,\cdots,x_{n-2})^T$, that is, multiplication by $A$ shifts each signal sample to the next vertex. 
Thus, $A$ is often called shift operator or shift matrix, which is similar to time shift operator $z^{-1}$ in DSP. In practice, adjacency matrix can be replaced by other matrices which reflect the structure of the graph, such as the Laplacian matrix and the normalized Laplacian matrix for undirected graphs, and the probability transition matrix. Here, we use $S$ to denote the general shift matrix, whether it is $A$, (normalized) Laplacian matrix, or the probability transition matrix.

In classical 1-D DSP, a shift-invariant filter $F$ 
has a $Z$-transform (polynomial representation in $z^{-1}$), that is
\begin{equation*}
F(z^{-1}) =\sum\limits_{k=-\infty}^{+\infty}f_k z^{-k},
\end{equation*}
where 
$f_k$ is polynomial coefficient. Moreover, from the shift-invariance property, it follows that the filtered output of a shifted input is equal to the shifted filtered output of the original input. In other words, the shift operation and the filter commute. That is,
$Fz^{-1}=z^{-1}F$, which directly follows from the above polynomial representation (see, e.g., \cite{Teke2017Linear}). 

Extending this concept to GSP, we also define a shift-invariant filter $H$ as the one that commutes with the shift matrix, i.e., $HS=SH$. 
However, unlike in the classical DSP case, a shift-invariant filter does not necessarily have a polynomial representation in terms of the shift operator $S$. Yet, $H$ can be represented as a polynomial in $S$ if the shift matrix $S$ satisfies the following condition.
 
\begin{myDef}[Shift-enabled graph{\cite{Liyanchen2018}}]
\label{def:shift_enable}
A graph ${\mathcal G}$ is shift-enabled if its corresponding shift matrix $S$ satisfies $p_S(\lambda) =m_S(\lambda)$, where $p_S(\lambda)$ and $m_S(\lambda)$ are the minimum polynomial and the characteristic polynomials of $S$, respectively. We also say that $S$ is shift-enabled when the above condition is satisfied. Otherwise, $S$ and the corresponding graph, are non-shift-enabled.
\end{myDef}

For shift-enabled graphs, the following theorem is the basis of linear, shift-invariant filter design.

\begin{Thm}{\label{Thm_shift_commute}}
The shift matrix $S$ is shift-enabled if and only if every matrix $H$ commuting with $S$ is a polynomial in $S$\rm~{\cite{sandryhaila_2013_discrete}}.
\end{Thm}

Note that this theorem implies that as long as the shift matrix $S$ does not satisfy the shift-enabled condition (i.e., $m_S(\lambda)\neq p_S(\lambda)$), there will always be some shift-invariant filters (and thus some filters) that cannot be represented as a polynomial of $S$. Ref~\cite{sandryhaila_2013_discrete} de-emphasized the shift-enabled condition by suggesting that we may work around it with the following theorem. 

\begin{Thm}[Theorem 2 in \cite{sandryhaila_2013_discrete}]\label{Theorem2_Moura}
For any shift matrix $S$, there exists a converted matrix $\tilde{S}$ and matrix polynomial $r(\cdot)$, such that $S=r(\tilde{S})$ and $m_{\tilde{S}}(\lambda)= p_{\tilde{S}}(\lambda)$.
\end{Thm}

While the above theorem is correct, it does not take into account that the target filter $H$ may not be shift-invariant with respect to the converted shift matrix. In particular, for a directed graph, in general, $S$ is not symmetric, and thus not jointly diagonalized with $H$. Consequently, one can show that generally there exists no converted shift-enabled $\tilde{S}$ that can maintain shift-invariance with the target filter 
 when the graph is directed and $S$ is asymmetric \cite{Liyanchen2018}. 

However, the conversion method suggested in \cite{sandryhaila_2013_discrete} does hold for undirected graphs when $H$ can be jointly diagonalized with $S$. 
Yet, as we will show in the following, the converted $\tilde S$ may not describe the same graph as the original $S$. This makes the whole conversion process moot. Hence, the shift-enabled condition is important regardless of whether the graph is directed or not (i.e., the shift matrix is asymmetric or not).




\section{The necessity of Shift-enabled condition for undirected graphs}

\label{sect:counterexample}
Before giving a concrete example, let us first review the conversion process described in \cite{sandryhaila_2013_discrete}. As mentioned earlier, even though the conversion process does not hold for arbitrary shift matrices, it can be applied to symmetric shift matrices.


According to Lemma~\ref{Lem_diag} in Appendix A, 
two symmetric and commuting matrices $S$ and $H$
are simultaneously diagonalizable. Thus, there exists an invertible matrix $T$ such that $S=T\Lambda_{S} T^{-1}$ and $H=T\Lambda_{H} T^{-1}$, where $\Lambda_{S}$ and $\Lambda_{H}$ are composed of the eigenvalues of $S$ and $H$, respectively. Then, a new matrix $\Lambda_{perturb}$ with distinct diagonal elements can be generated by slightly perturbing the values of $\Lambda_{S}$. The new shift matrix is calculated as $\tilde{S}=T\Lambda_{perturb} T^{-1}$.
According to Lemma~\ref{Lem_undirected} and Lemma~\ref{Lem_diag}, the restructured shift matrix $\tilde{S}$ satisfies
$p_{\tilde{S}}(\lambda)=m_{\tilde{S}}(\lambda)$ and $H\tilde{S}=\tilde{S}H$.
Hence, from Theorem~\ref{Thm_shift_commute}, $H$ is a polynomial in $\tilde{S}$. 

However, it is not sufficient to have $H$ represented as a polynomial of any arbitrary $\tilde{S}$. 
A natural and basic constraint is that
the converted $\tilde{S}$ should facilitate ``local processing'', that is, it should describe topologically the same graph, which is essential in virtually all GSP applications, such as filter design~\cite{Hammond2009Wavelets},
sampling~\cite{Gadde2014Sampling},  denoising~\cite{ChengYang2017Estimating}, and classification~\cite{HeKangHang2018Non}, otherwise, the conversion is meaningless. To ensure that the converted graph facilitates ``local processing'', that is, an implementation of an $L$-th order polynomial filter requires $L$ data exchanges between neighbouring nodes~\cite{Teke2017Linear}, we introduce Definition~\ref{same_structure}.
In fact, the definition of a matrix describing a graph (see details in Definition~\ref{same_structure}) is not new in 
spectral graph theory. In
particular, the matrix of ``loose description'' is widely used in the context of inverse eigenvalue
problem and zero-forcing problem~\cite{hogben2005spectral,trefois2015zero}. We introduce ``strict description'' since we would
like to accommodate graphs with self-loops.
In a nutshell, two shift matrices describe the same graph if the conversion from one to another preserves the graph topological structure, implying that filtering under the converted graph can be performed locally. The precise definition is specified as follows.
\begin{myDef}[\cite{hogben2005spectral,fallat2007minimum,trefois2015zero}]
{\label{same_structure}}
Shift matrices $S$ and $\tilde{S}$  strictly describe the same graph if 
1) $S_{i,j}\neq 0$ if and only if $\tilde{S}_{i,j}\neq 0$ for any $i$ and $j$, and 2) $\tilde{S}$ is symmetric if and only if $S$ is symmetric. And we will say $S$ and $\tilde{S}$ loosely describe the same graph if the first condition is relaxed to 1') $S_{i,j}\neq 0$ if and only if $\tilde{S}_{i,j}\neq 0$ only for $i\neq j$. That is, we allow some $i$ where only $S_{i,i}$ or $\tilde{S}_{i,i}$ equal to $0$. 
\end{myDef}

Given this additional constraint that $\tilde S$ and $S$ should describe the same graph structure,
we can show that it is impossible to guarantee the following three conditions to be satisfied simultaneously:
\begin{itemize}
\item
 $\tilde{S}$ is shift-enabled (i.e., $p_{\tilde{S}}(\lambda)=m_{\tilde{S}}(\lambda)$).
 \item $H$ is shift-invariant on $\tilde{S}$ (i.e.,  $H\tilde{S}=\tilde{S}H$).
 \item
 $\tilde{S}$ and $S$ strictly or loosely describe the same graph.
\end{itemize}




\subsection{A counter-example that $\tilde{S}$ can loosely but not strictly describe the original graph
}
\begin{figure}[t]
        \centering
        \subfigure[]{
        \label{figa}
     \includegraphics[width=1.05in]{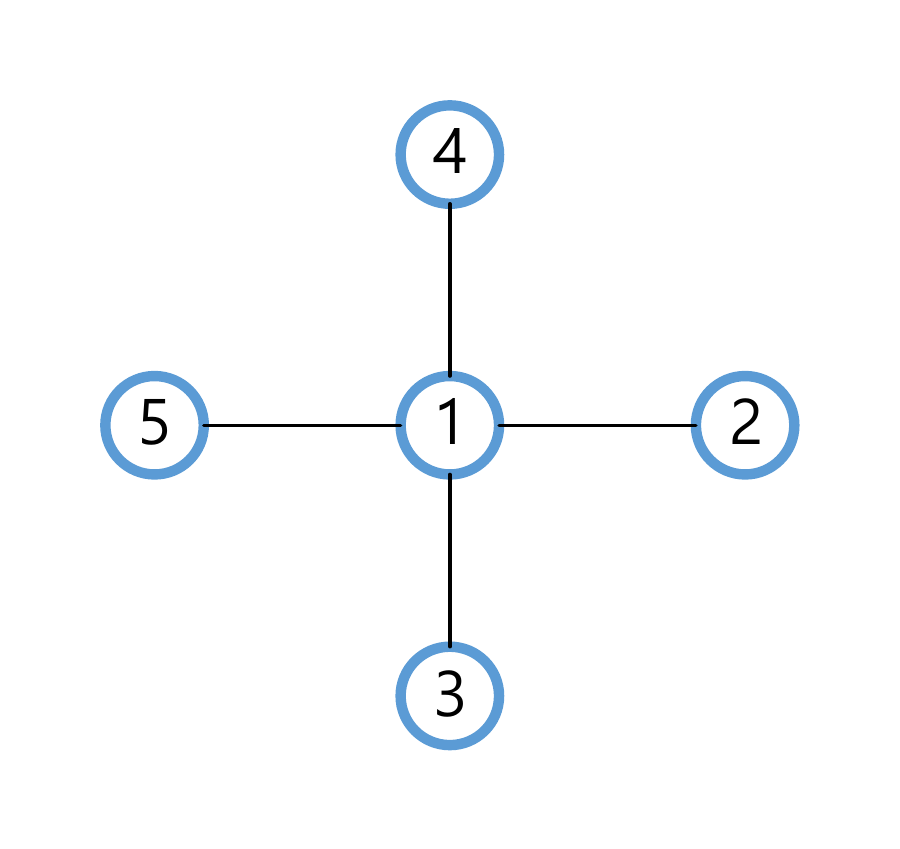}}
    \centering
        \subfigure[]{
         \label{figb} 
 \includegraphics[width=1.05in]{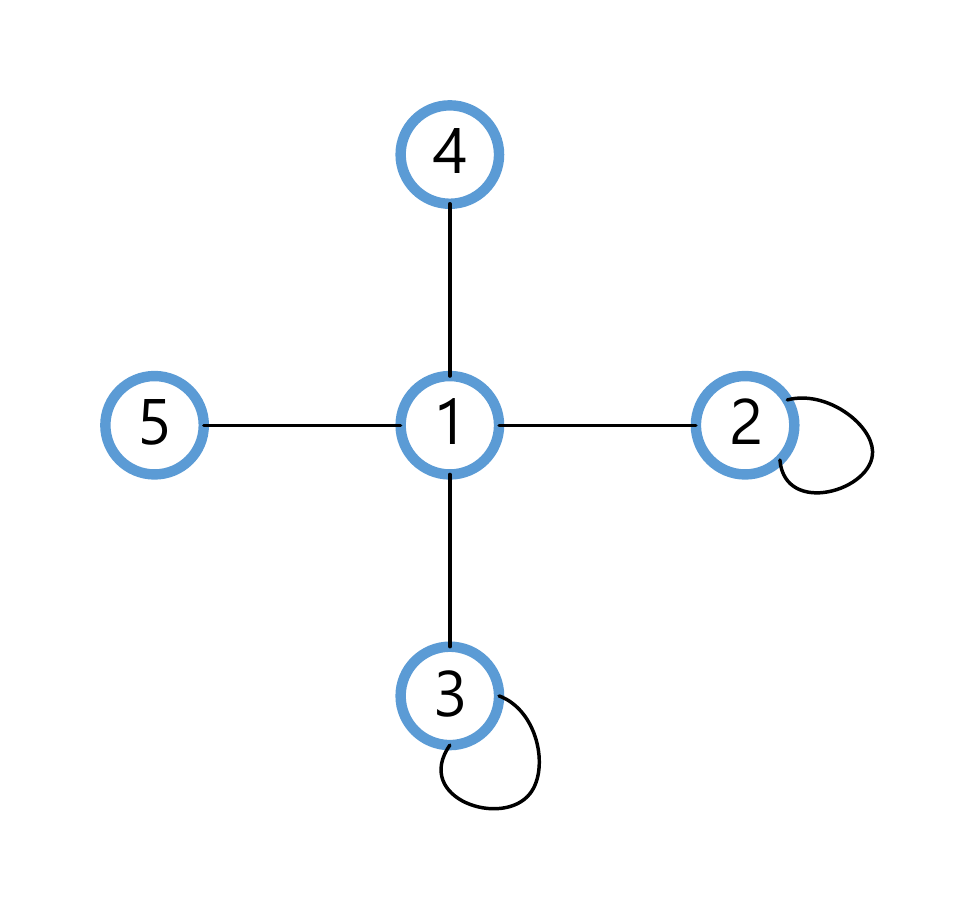}}
  \centering
        \subfigure[]{
        \label{figcycle}
        \includegraphics[width=0.85in]{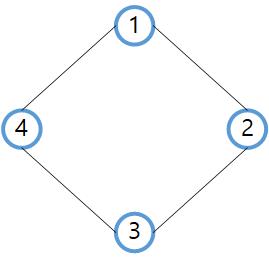}}
        \caption{Graph topology used in the examples. (a) Original graph with shift matrix $S$. (b) Converted shift matrix $\tilde{S}$ which loosely describes the same graph as $S$. (c) Cycle graph with shift matrix $S'$. 
}
\end{figure}\label{Fig3}

Let us start with a non-shift-enabled graph
as shown in Figure.~\ref{figa}. The shift matrix\footnote{Without loss of generality, we choose adjacency matrix as the shift matrix in the following examples.} of the undirected graph is
$S=\begin{pmatrix}
\begin{smallmatrix}
0 & 1 & 1 & 1 & 1\\
1 & 0 & 0 & 0 & 0\\
1 & 0 & 0 & 0 & 0\\
1 & 0 & 0 & 0 & 0\\
1 & 0 & 0 & 0 & 0\\
\end{smallmatrix}
\end{pmatrix}. 
$ 
It is clear that 
$p_S\left( \lambda \right)=\lambda^3\left( \lambda-2\right) \left(\lambda +2 \right)\neq \lambda\left( \lambda-2\right) \left(\lambda +2 \right)=m_S(\lambda)$
and hence $S$ is non-shift-enabled.
Since 
shift-enabled condition is not just sufficient but also necessary~\cite{Liyanchen2018}, there must exist a shift-invariant filter not representable as a polynomial of $S$. 
Indeed, one example for such a filter is 
\vspace{5pt}
$H=\begin{pmatrix}
\begin{smallmatrix}
0 & 0 & 0 & 0 & 0\\
0 & 1 & -1 & 0 & 0\\
0 & -1 & 1 & 0 & 0\\
0 & 0 & 0 & 0 & 0\\
0 & 0 & 0 & 0 & 0\\
\end{smallmatrix}
\end{pmatrix} 
$. It can be readily verified that $HS=0=SH$ and thus the filter is shift-invariant, and it is impossible to find polynomial representation of $H$ in terms of $S$. Note that
$S^{n}_{2,3}=S^{n}_{2,4}$\footnote{Note that $S_{i,j}^k$ denotes the $(i,j)$-element of matrix $S^k$.}  for all $n\in\mathbb{N}$.
Thus for any polynomial $h(S)$, we must have $h(S)_{2,3}=h(S)_{2,4}$. 
But since $H_{2,3}=-1\neq 0=H_{2,4}$,  $H\neq h\left ( S \right) $ for any polynomial function $h(\cdot)$. 

\subsubsection{Extension of $H$ to a class of filters}
Note that we can extend $H$ to the following class of filters that all cannot be represented as polynomials of $S$:
\begin{equation}
\mathbb{H}=\{\alpha H+ q(S)|\alpha \in \mathbb{R}, q(S) \mbox{ is a polynomial of $S$}\} \label{a_class_of_H}.
\end{equation}

Since apparently $q(S)S=Sq(S)$ for any polynomial $q(S)$ and $HS=SH$ as discussed above, any filter $\alpha H + q(S) \in \mathbb{H}$ commutes with $S$ as well.
Thus any filter in $\mathbb{H}$ is shift-invariant. However, since $H$ is not representable as a polynomial of $S$, as discussed above, so does $\alpha H +q(S)$.

From the examples presented above, we note that when the shift-enabled condition 
is violated, we may find an infinite number of shift-invariant filters that are not representable as polynomials of $S$. 

\subsubsection{Shift-enabled $\tilde S$ that strictly describes the original graph does not exist}
First, let us restrict the converted shift matrix $\tilde{S}$ to strictly describe the same graph as $S$. 
Thus $\tilde{S}$ could be written as
\begin{equation}
\tilde{S}=\begin{pmatrix}
\begin{smallmatrix}
0 & {\tilde{S}}_{1,2} & {\tilde{S}}_{1,3} & {\tilde{S}}_{1,4} & {\tilde{S}}_{1,5}\\ 
{\tilde{S}}_{1,2} & 0 & 0 & 0 & 0\\
{\tilde{S}}_{1,3} & 0 & 0 & 0 & 0\\
{\tilde{S}}_{1,4} & 0 & 0 & 0 & 0\\
{\tilde{S}}_{1,5} & 0 & 0 & 0 & 0\\
\end{smallmatrix}
\end{pmatrix}
\end{equation}
with non-zeros $\tilde{S}_{1,2}, \tilde{S}_{1,3}, \tilde{S}_{1,4}$, and $\tilde{S}_{1,5}$.
We can readily verify that the characteristic polynomial is 
$p_{\tilde{S}}(\lambda)=\lambda^3(\lambda^2-{\tilde{S}}_{12}^2-{\tilde{S}}_{13}^2-{\tilde{S}}_{14}^2-{\tilde{S}}_{15}^2)$ and 0 is the triple eigenvalue of $\tilde{S}$. According to Lemma~\ref{Lem_undirected}, a  shift-enabled real symmetric shift matrix has to have unique eigenvalues and thus ${\tilde{S}}$ is not shift-enabled. Therefore, all graphs which have the
same structure as Figure~\ref{figa} are non-shift-enabled. 


\subsubsection{Shift-enabled $\tilde{S}$ that loosely describes the original graph exists}
Next, let us relax $\tilde{S}$ so that it may just loosely describe the original graph. In other words, we allow the diagonal elements to be non-zero which maintains most of the topological structure of the original graph. In applications where 
diffusion or state transition matrices are treated as shift matrices, the diagonal elements can be interpreted as the returning probabilities of  the current state to itself. 
Thus, the converted shift matrix $\tilde S$ can be written as
\begin{equation}
\tilde{S}=\begin{pmatrix}
\begin{smallmatrix}
{\tilde{S}}_{1,1} & {\tilde{S}}_{1,2} & {\tilde{S}}_{1,3} & {\tilde{S}}_{1,4} & {\tilde{S}}_{1,5}\\
{\tilde{S}}_{1,2} & {\tilde{S}}_{2,2} & 0 & 0 & 0\\
{\tilde{S}}_{1,3} & 0 & {\tilde{S}}_{3,3} & 0 & 0\\
{\tilde{S}}_{1,4} & 0 & 0 & {\tilde{S}}_{4,4} & 0\\
{\tilde{S}}_{1,5} & 0 & 0 & 0 & {\tilde{S}}_{5,5}\\
\end{smallmatrix}
\end{pmatrix}. 
\end{equation}

Many solutions that satisfy shift-enabled and shift-invariant conditions can be found. 
For instance,   
$\tilde{S}=\begin{pmatrix}
\begin{smallmatrix}
0 & 1& 1 & 1 & 1\\
1 & 1 & 0 & 0 & 0\\
1 & 0 & 1 & 0 & 0\\
1 & 0 & 0 & 0 & 0\\
1 & 0 & 0 & 0 & 0\\
\end{smallmatrix}
\end{pmatrix}$ is one such solution, where the original graph structure is only slightly modified as shown in Figure~\ref{figb}. One can verify that the eigenvalues $(-1.8136,0,0.4707,1,2.3429)$ of $\tilde{S}$ are distinct and thus $\tilde S$ is shift-enabled.  Moreover, one can also readily verify that  $H\tilde{S}=\tilde{S}H$. By Theorem 1, the above two conditions ensure that $H$ is a polynomial in $\tilde{S}$. 

\subsection{A counter example when
the converted shift matrix
can neither strictly nor loosely describe the original graph}

\label{sect:example2}

Note that there are situations where no shift-enabled $\tilde S$ exists even after we relax the graph structure constraint as in the earlier example. 
Consider shift matrix $S'=
\begin{pmatrix}
\begin{smallmatrix}
0 & 1 & 0 & 1 \\
1 & 0 & 1 & 0\\
0 & 1 & 0 & 1 \\
1 & 0 & 1 & 0\\
\end{smallmatrix}
\end{pmatrix} 
$ as shown in Figure~\ref{figcycle}. 



It can easily be seen that the eigenvalues of $S'$, $(0,0,2,-2)$, are not unique. 
Thus $S'$ is non-shift-enabled according to Lemma~\ref{Lem_undirected}. 
So we do expect that there exists shift-invariant filter not representable by $S'$. Indeed, we can easily show that filter $H'=\begin{pmatrix}
\begin{smallmatrix}
0 & 0 & -1 & 1 \\
0 & -1 & 1 & 0\\
-1 & 1 & 0 & 0 \\
1 & 0 & 0 & -1\\
\end{smallmatrix}
\end{pmatrix} 
$ is such a filter. 

First, note that $H' S'=S' H'$ and thus $H'$ is shift-invariant under $S'$.
Furthermore, note that 
$(S')^{n}_{1,2}=(S')^{n}_{1,4}$ for all $n\in\mathbb{N}$, and so $h(S')_{1,2}=h(S')_{1,4}$ for any polynomial $h({S'})$.  But since $H'_{1,2}=0\neq 1=H'_{1,4}$,  $H'\neq h\left ( S' \right) $ for any polynomial function $h(\cdot)$.

Let us prove that it is impossible to find a converted shift matrix $\tilde{S'}$ which is shift-enabled and commutes with $H'$ by only changing the weights of nonzero and diagonal elements.

Consider a general symmetric matrix
\begin{equation}\label{Equ_5}
\tilde{S'}=\begin{pmatrix}
\begin{smallmatrix}
\tilde{S'}_{1,1} & \tilde{S'}_{1,2} & 0 & \tilde{S'}_{1,4} \\
\tilde{S'}_{1,2} & \tilde{S'}_{2,2} & \tilde{S'}_{2,3} & 0\\
0 & \tilde{S'}_{2,3} & \tilde{S'}_{3,3} &\tilde{S'}_{3,4} \\
\tilde{S'}_{1,4} & 0 & \tilde{S'}_{3,4} & \tilde{S'}_{4,4}\\
\end{smallmatrix}
\end{pmatrix} 
\end{equation}
which has arbitrary weights on nonzero and diagonal elements. That is, $\tilde{S'}$ loosely describes the same graph as $S'$.

$H'=h(\tilde{S'})$ clearly implies that $H'$ commutes with $\tilde{S'}$, namely, $H'\tilde{S'}=\tilde{S'}H'$ is a necessary condition for $H'=h(\tilde{S'})$.
It follows from $H'\tilde{S'}=\tilde{S'}H'$ that $\tilde{S'}_{1,1}=\tilde{S'}_{2,2}=\tilde{S'}_{3,3}=\tilde{S'}_{4,4}$ and $\tilde{S'}_{1,2}=\tilde{S'}_{1,4}=\tilde{S'}_{2,3}=\tilde{S'}_{3,4}$
, i.e., 
\begin{equation}\label{Equ_6}
\tilde{S'}=\begin{pmatrix}
\begin{smallmatrix}
\tilde{S'}_{1,1} & \tilde{S'}_{1,2} & 0 & \tilde{S'}_{1,2} \\
\tilde{S'}_{1,2} & 
\tilde{S'}_{1,1} & 
\tilde{S'}_{1,2} & 0\\
0 & \tilde{S'}_{1,2} & \tilde{S'}_{1,1} & \tilde{S'}_{1,2} \\
\tilde{S'}_{1,2} & 0 &\tilde{S'}_{1,2} & \tilde{S'}_{1,1}\\
\end{smallmatrix}
\end{pmatrix}. 
\end{equation}

Following Cayley-Hamilton Theorem \cite{lancaster_1985_matrix_theory}, if $H'$ is a polynomial in $\tilde{S'}$, then $H'=h(\tilde{S'})=h_0I+h_1\tilde{S'}+h_2\tilde{S'}^2+h_3\tilde{S'}^3$, where $I$ as the identity matrix. 
In fact, it is easy to prove that $(\tilde{S'})^k_{1,2}=(\tilde{S'})^k_{1,4}$, for $k=0,1,2,3$. Hence,  $h(\tilde{S'})_{1,2}=h(\tilde{S'})_{1,4}$ which contradicts with $H'_{1,2}\neq H'_{1,4}$. Thus, for this example, the filter $H'$ cannot be represented as a polynomial in the converted shift matrix $\tilde{S'}$ which even just loosely describes the original graph.

\section{Conclusion}
For a non-shift-enabled graph, even if
we can easily ``transform" the symmetric shift matrix $S$ into one that satisfies the shift-enabled condition, the new $\tilde{S}$ may be irrelevant since it describes a very different graph from $S$. 
That is, the operator $\tilde{S}$ on a graph signal may involve mixing inputs far beyond its neighborhood and become impractical for huge graphs.
Combined with the necessity of the shift-enabled condition for directed graph~\cite{Liyanchen2018}, we demonstrated in this letter that the shift-enabled condition is essential for any graph structure. 
{\color{blue}
}

Note that even though we consider the adjacency matrix as the shift matrix in our examples, the conclusion applies to other shift matrices. In particular, one can readily verify that the conclusion still holds if we use the Laplacian matrix as the shift matrix in the example in Section \ref{sect:example2}.

\appendices
\section{} 
It is easily determined whether a graph is shift-enabled by the following lemmas.



\begin{lemma}{\label{Lem_undirected}}
	If shift matrix $S$ is a real symmetric matrix, then $S$ is shift-enabled, if and only if all eigenvalues of $S$ are distinct {\rm~\cite{ortega2018graph}}.
\end{lemma}

Lemma~\ref{Lem_undirected} indicates that an undirected graph is shift-enabled if and only if its eigenvalues are all distinct.

As both shift matrix $S$ and filter matrix $H$ are symmetric, we can obtain the following lemma.

\begin{lemma}{\label{Lem_diag}}
If shift matrix $S$ and filter matrix $H$ are diagonalizable (this condition always holds for symmetric matrix) then $S$ and $H$ are simultaneously diagonalizable (by an invertible matrix) if and only if $HS=SH$~\rm{(see Theorem 1.3.12 in \cite{horn_2012_matrix_analysis})}.
\end{lemma}


\section*{Acknowledgment}
The authors thank B. Zhao for helpful discussions. This project has received funding from  the European Union`s Horizon 2020 research and innovation programme under the Marie Sklodowska-Curie grant agreement no. 734331 and  the Fundamental Research Funds for the Central Universities no. 0800219369.

\ifCLASSOPTIONcaptionsoff
  \newpage
\fi




\bibliographystyle{IEEEtran}
\bibliography{reference}

\end{document}